\documentclass[reqno]{amsart}
\usepackage{amssymb}
\usepackage{amsmath}
\usepackage{amsfonts}

\newtheorem{theorem}{Theorem}
\theoremstyle{plain}

\newtheorem{corollary}{Corollary}

\newtheorem{definition}{Definition}
\newtheorem{example}{Example}

\newtheorem{lemma}{Lemma}

\newtheorem{proposition}{Proposition}

\numberwithin{equation}{section}
\input{tcilatex}

\begin{document}
\title[$p$-adic Parabolic Equations and Ultrametric Diffusion]{Taibleson
Operators, $p$-adic Parabolic Equations and Ultrametric Diffusion.}
\author{J. J. Rodr\'{\i}guez-Vega}
\address{Departamento de Matem\'{a}ticas, Universidad Nacional de Colombia,
Ciudad Universitaria, Bogot\'{a} D.C., Colombia.}
\email{jjrodriguezv@unal.edu.co}
\author{W. A. Z\'{u}\~{n}iga-Galindo}
\address{Centro de Investigaci\'{o}n y de Estudios Avanzados del I.P.N.,
Departamento de Matem\'{a}ticas, Av. Instituto Polit\'{e}cnico Nacional
2508, Col. San Pedro Zacatenco, M\'{e}xico D.F., C.P. 07360, M\'{e}xico. }
\email{wzuniga@math.cinvestav.mx}
\subjclass[2000]{Primary 35R60, 60J25; Secondary 47S10, 35S99}
\keywords{Parabolic equations, Markov processes, $p$-adic numbers,
ultrametric diffusion.}

\begin{abstract}
We give a multimensional version of the p-adic heat equation, and show that
its fundamental solution is the transition density of a Markov process.
\end{abstract}

\maketitle

\section{Introduction}

In recent years $p-$adic analysis has received a lot of attention due to its
applications in mathematical physics, see e.g. \cite{A-K1}, \cite{A-K2}, 
\cite{A-B-K-O}, \cite{A-B-O}, \cite{Kh1}, \cite{Kh2}, \cite{Koch1}, \cite%
{R-T}, \cite{VVZ}\ and references therein. One motivation comes from
statistical \ physics, \ in particular in connection with models describing
relaxation in glasses, macromolecules, and proteins. It has been proposed
that the non exponential nature of those relaxations is a consequence of a
hierarchical structure of the state space which \ can in turn \ be put in
connection with $p-$adic structures (\cite{A-B-K-O}, \cite{A-B-O}, \cite{R-T}%
). In \cite{A-B-K-O} was demostrated that the $p$-adic analysis is a natural
basis for the construction of a wide variety of models of ultrametric
diffusion constrained by hierarchical energy landscapes. To each of these
models\ is associated a stochastic equation (the master equation). In
several cases this equation is a $p$-adic parabolic equation of type:

\begin{equation}
\begin{cases}
\frac{\partial u(x,t)}{\partial t}+a(Au)(x,t)=f(x,t),\quad x\in \mathbb{Q}%
_{p}^{n},\quad t\in (0,T], \\ 
\\ 
u(x,0)=\varphi (x),%
\end{cases}
\label{1}
\end{equation}%
where $a$ is a positive constant, $A$ is pseudo-differential operator, and $%
\mathbb{Q}_{p}$ is the field of $p$-adic numbers. The simplest case occurs
when $n=1$ and $A$ is the Vladimirov operator:%
\begin{equation*}
\left( D^{\alpha }\varphi \right) (x)=\mathcal{F}_{\xi \rightarrow
x}^{-1}\left( \left\vert \xi \right\vert _{p}^{\alpha }\mathcal{F}%
_{x\rightarrow \xi }\varphi (x)\right) ,\text{ }\alpha >0,
\end{equation*}%
where $\mathcal{F}$ is the Fourier transform. The fundamental solution of (%
\ref{1}) is density transition of a time- and space-homogeneous Markov
process, that is consider the $p-$adic counterpart of the Brownian motion
(see \cite{Koch1}, \cite{VVZ}).

It is relevant to mention that in the case $n=1$, the fundamental solution
of (\ref{1}) \ when $A=D^{\alpha }$ (also called the $p-$adic heat kernel)
has been studied extensively, see e.g. \cite{Bla}, \cite{Ha1}, \cite{Ha2}, 
\cite{I}, \cite{Koch1}, \cite{VVZ}.

A natural problem is to study the initial value problem (\ref{1}) in the $n$%
-dimensional case. Recently, the second author considered Cauchy's problem (%
\ref{1}) when 
\begin{equation*}
\left( A\varphi \right) \left( x\right) =\mathcal{F}_{\xi \rightarrow
x}^{-1}\left( \left\vert f\left( \xi \right) \right\vert _{p}^{\alpha }%
\mathcal{F}_{x\rightarrow \xi }\varphi (x)\right) ,\text{ }\alpha >0,
\end{equation*}%
here $f\left( \xi \right) $ is an elliptic homogeneous polynomial in $n$
variables, and the datum $\varphi $ is a locally constant and integrable
function. Under these hypotheses it was established the existence of a
unique solution to Cauchy's problem (\ref{1}). In addition, the fundamental
solution is a transition density of a Markov process \ with space state $%
\mathbb{Q}_{p}^{n}$ (see \cite{Z10}).

In this paper we study Cauchy's problem (\ref{1}) when $A$ is the Taibleson
pseudo-differential operator which is defined as follows:%
\begin{equation}
\left( D_{T}^{\beta }\varphi \right) \left( x\right) =\mathcal{F}_{\xi
\rightarrow x}^{-1}\left( \left( \max_{1\leq i\leq n}\left\vert \xi
_{i}\right\vert _{p}\right) ^{\beta }\mathcal{F}_{x\rightarrow \xi }\varphi
(x)\right) ,\text{ }\beta >0.  \label{4}
\end{equation}%
Recently Albeverio, Khrennikov, and Shelkovich studied $D_{T}^{\beta }$ in
the context of the Lizorkin spaces \cite{A-K-S}.

We prove \ existence and uniqueness of the Cauchy problem (\ref{1}-\ref{4})
in spaces of increasing functions introduced by Kochubei in \cite{Koch2},
see Theorem \ref{Thm 1}. We also associate a Markov processes to equation
the fundamental solution (see Theorem \ref{theo3}). These results constitute
an extension of the corresponding results in \cite{Koch1}, \cite{VVZ}.

We want to mention here a relevant comment due to the referee. There exists
a procedure, developed in \cite{Koch1} for elliptic equations, of reducing
multi-dimensional problems over $\mathbb{Q}_{p}$ to one-dimensional problems
over appropriate field extensions. In particular, the Taibleson operator is
connected with the unramified extension of $\mathbb{Q}_{p}$ of degree $n$
(see Lemma 2.1 in \cite{Koch1}). The fundamental solutions corresponding to
the multi-dimensional Cauchy problem and the problem over the unramified
extension should be obtained from each other, up to a linear change of
variables, as in the formula (2.38) of \cite{Koch1} for the elliptic case.
Then many properties of the fundamental solution would follow directly from
those known in the one-dimensional case. In this paper we use an elementary
and independent method that has its obvious advantages.

Let us explain the connection between the results of this paper and those of 
\cite{Z10}. There are infinitely many homogeneous polynomial functions
satisfying 
\begin{equation*}
\left\vert f\left( \xi \right) \right\vert _{p}=\left( \max_{1\leq i\leq
n}\left\vert \xi _{i}\right\vert _{p}\right) ^{d},\text{ for any }\xi \in 
\mathbb{Q}_{p}^{n},
\end{equation*}%
here $d$ denotes degree of $f$ (c.f. Lemmas \ref{lemma15}-\ref{lemma16}).
Hence the pseudo-differential operators considered here are a subclass of
the ones considered in \cite{Z10}. However, the function spaces for the
solutions and initial data\ are completely different. In this paper the
initial datum and the solution to Cauchy problem (\ref{1}-\ref{4}) are not
necessarily bounded, neither integrable, but in \cite{Z10} are.

Finally, our results can be extended to operators of the form%
\begin{equation}
\left( A\varphi \right) \left( x\right) =a_{0}(x,t)(D_{T}^{\alpha }\varphi
)(x)+\sum_{k=1}^{n}a_{k}(x,t)(D_{T}^{\alpha _{k}}\varphi )(x)+b(x,t)\varphi
(x),  \label{3}
\end{equation}%
$\alpha >1$, $0<\alpha _{1}<\dotsc <\alpha _{n}<\alpha $, where the $%
a_{k}(x,t)$ ,and $b(x,t)$ are bounded continuous functions, using the
techniques presented in \cite{Koch1}-\cite{Koch3}. These results will appear
later elsewhere.

The authors wish to thank the referee for the relevant comment mentioned
above.

\section{Preliminary Results}

As general reference for $p$-adic\ analysis we refer the reader to \cite{TA}%
\ and \cite{VVZ}. The field of $p$-adic numbers $\mathbb{Q}_{p}$ is defined
as the completion of the field of rational numbers $\mathbb{Q}$ with respect
to the non-Archimedean $p$-adic norm $|\cdot |_{p}$ which is defined as
follows: $|0|_{p}=0$; if $x\in \mathbb{Q}^{\times }$, $x=p^{\gamma }\frac{a}{%
b}$ with $a$, $b$ integers coprime to $p$, then $\left\vert x\right\vert
_{p}=p^{-\gamma }$. The integer $\gamma =\gamma \left( x\right) $ is called
the $p$\textit{-adic order of} $x$, and it will be denoted as $ord\left(
x\right) $. We use the same symbol, $|\cdot |_{p}$, for the $p$-adic norm on 
$\mathbb{Q}_{p}$. We extend the $p$-adic norm to $\mathbb{Q}_{p}^{n}$ as
follows:

\begin{equation*}
\left\Vert x\right\Vert _{p}:=\max_{1\leq i\leq n}\left\vert
x_{i}\right\vert _{p}\text{, for }x=\left( x_{1},\ldots ,x_{n}\right) \in 
\mathbb{Q}_{p}^{n}\text{.}
\end{equation*}%
Note that $\left\Vert x\right\Vert _{p}=p^{-\min_{1\leq i\leq n}\left\{
ord\left( x_{i}\right) \right\} }$.

Any $p$-adic number $x\neq 0$ has a unique expansion of the form 
\begin{equation*}
x=p^{\gamma }\dsum\limits_{j=0}^{\infty }x_{j}p^{j},
\end{equation*}%
where $\gamma =ord\left( x\right) \in \mathbb{Z}$, and $x_{j}\in \left\{
0,1,\ldots ,p-1\right\} $. By using the above expansion, we define the 
\textit{fractional part of }$x\in \mathbb{Q}_{p}$, denoted as $\left\{
x\right\} _{p}$, as the following rational number: 
\begin{equation*}
\left\{ x\right\} _{p}:=\left\{ 
\begin{array}{llll}
0, & \text{if} & x=0\text{,} & \text{ or \ }\gamma \geq 0 \\ 
p^{\gamma }\dsum\limits_{j=0}^{\left\vert \gamma \right\vert -1}x_{j}p^{j},
& \text{if} & \gamma <0. & 
\end{array}%
\right.
\end{equation*}%
Denote by $B_{\gamma }^{n}\left( a\right) =\left\{ x\in \mathbb{Q}%
_{p}^{n}\mid \left\Vert x-a\right\Vert _{p}\leq p^{\gamma }\right\} $, the
ball of radius $p^{\gamma }$ with center at $a=\left( a_{1},\ldots
,a_{n}\right) \in \mathbb{Q}_{p}^{n}$, and $B_{\gamma }^{n}\left( 0\right)
=B_{\gamma }^{n}$, $\gamma \in \mathbb{Z}$. Note that $B_{\gamma }^{n}\left(
a\right) =B_{\gamma }\left( a_{1}\right) \times \ldots \times B_{\gamma
}\left( a_{n}\right) $, where $B_{\gamma }\left( a_{j}\right) =\left\{
x_{j}\in \mathbb{Q}_{p}\mid \left\vert x_{j}-a_{j}\right\vert _{p}\leq
p^{\gamma }\right\} $ is the one-dimensional ball of radius $p^{\gamma }$
with center at $a_{j}\in \mathbb{Q}_{p}$. The \ Ball $B_{0}^{n}$ equals the
product of n copies of $B_{0}\left( 0\right) =\mathbb{Z}_{p}$, the ring of $%
p $-adic integers.

Let $d^{n}x$ denote the Haar measure on $\mathbb{Q}_{p}^{n}$ normalized by
the condition $\tint\nolimits_{B_{0}^{n}}d^{n}x=1$.

A complex-valued function $\varphi$\ defined on $\mathbb{Q}_{p}^{n}$ is
called \textit{locally constant} if for any $x\in\mathbb{Q}_{p}^{n}$ there
exists an integer $l(x)\in\mathbb{Z}$ such that $\varphi\left(
x+x^{\prime}\right) =\varphi\left( x\right) $, for $x^{\prime}\in
B_{l(x)}^{n}$.

A function $\varphi :\mathbb{Q}_{p}^{n}\rightarrow \mathbb{C}$ is called 
\textit{Schwartz-Bruhat function}, or \textit{test function}, if it is%
\textit{\ }locally constant with compact support. The $\mathbb{C}$-vector
space of the Schwartz-Bruhat functions is denoted by $S(\mathbb{Q}_{p}^{n})$%
. If $\varphi \in S(\mathbb{Q}_{p}^{n})$, there exist an integer $l\geq 0$
such that $\varphi \left( x+x^{\prime }\right) =\varphi \left( x\right) $,
for $x^{\prime }\in B_{-l}^{n}$, and $x\in \mathbb{Q}_{p}^{n}$ (see e.g. 
\cite[VI.1, Lemma 1]{VVZ}). The largest of such numbers $l=l\left( \varphi
\right) $ is called \ \textit{the exponent of local constancy of} $\varphi $.

Let $S^{\prime }(\mathbb{Q}_{p}^{n})$ denote the \ set of all functionals
(distributions) on $S(\mathbb{Q}_{p}^{n})$. All the functionals on $S(%
\mathbb{Q}_{p}^{n})$ are continuous (see e.g. \cite[VI.3]{VVZ}).

Given $\xi=\left( \xi_{1},\ldots,\xi_{n}\right) $, $x=\left(
x_{1},\ldots,x_{n}\right) \in\mathbb{Q}_{p}^{n}$, we set $\xi\cdot x:=\tsum
\nolimits_{i=1}^{n}\xi_{i}x_{i}$. The Fourier transform \ of $\varphi\in S(%
\mathbb{Q}_{p}^{n})$ is defined as 
\begin{equation*}
(\mathcal{F}\varphi)(\xi)=\int_{\mathbb{Q}_{p}^{n}}\Psi(-\xi\cdot
x)\varphi(\xi)\,d^{n}x\text{, }\xi\in\mathbb{Q}_{p}^{n}\text{,}
\end{equation*}
where $\Psi(-\xi\cdot x)=\tprod
\nolimits_{i=1}^{n}\Psi(-\xi_{i}x_{i})=\exp\left( 2\pi i\tsum
\nolimits_{i=1}^{n}\left\{ -\xi_{i}x_{i}\right\} _{p}\right) $. The function 
$\Psi(\alpha x_{j})=\exp\left( 2\pi i\tsum \nolimits_{i=1}^{n}\left\{ \alpha
x_{j}\right\} _{p}\right) $ is called \textit{the standard additive character%
} of $\mathbb{Q}_{p}$. The Fourier Transform is a linear isomorphism from $S(%
\mathbb{Q}_{p}^{n})$ onto itself.

\subsection{The Taibleson Operator}

We set 
\begin{equation*}
\Gamma _{p}^{\left( n\right) }(\alpha ):=\frac{1-p^{\alpha -n}}{1-p^{-\alpha
}}\text{, }\alpha \neq 0.
\end{equation*}%
This function is called the $p$\textit{-adic Gamma function}. The function 
\begin{equation*}
\mathit{k}_{\alpha }(x)=\frac{||x||_{p}^{\alpha -n}}{\Gamma _{p}^{\left(
n\right) }\left( \alpha \right) },\quad \alpha \in \mathbb{R\setminus }%
\left\{ 0,n\right\} ,\quad x\in \mathbb{Q}_{p}^{n},
\end{equation*}%
is called \textit{the multi-dimensional Riesz Kernel}; it determines a
distribution on $S(\mathbb{Q}_{p}^{n})$ as follows. If $\alpha \neq 0$, $n$,
and $\varphi \in S(\mathbb{Q}_{p}^{n})$, then 
\begin{equation*}
\left\langle \mathit{k}_{\alpha }(x),\varphi (x)\right\rangle =\frac{1-p^{-n}%
}{1-p^{\alpha -n}}\varphi (0)+\frac{1-p^{-\alpha }}{1-p^{\alpha -n}}%
\int_{||x||_{p}>1}||x||_{p}^{\alpha -n}\varphi (x)\,d^{n}x
\end{equation*}%
\begin{equation}
+\frac{1-p^{-\alpha }}{1-p^{\alpha -n}}\int_{||x||_{p}\leq
1}||x||_{p}^{\alpha -n}(\varphi (x)-\varphi (0))\,d^{n}x.  \label{5}
\end{equation}%
Then $\mathit{k}_{\alpha }\in S^{\prime }(\mathbb{Q}_{p}^{n})$, for $\mathbb{%
R\setminus }\left\{ 0,n\right\} $. In the case $\alpha =0$, by passing to
the limit in (\ref{5}), we obtain 
\begin{equation*}
\langle \mathit{k}_{0}(x),\varphi (x)\rangle :=\lim_{\alpha \rightarrow
0}\left\langle \mathit{k}_{\alpha }(x),\varphi (x)\right\rangle =\varphi (0),
\end{equation*}%
i.e., \ $\mathit{k}_{0}(x)=\delta \left( x\right) $, the Dirac delta
function, and therefore $\mathit{k}_{\alpha }\in S^{\prime }(\mathbb{Q}%
_{p}^{n})$, for $\mathbb{R\setminus }\left\{ n\right\} $.

It follows from (\ref{5}) that for $\alpha >0$, 
\begin{equation}
\langle \mathit{k}_{-\alpha }(x),\varphi (x)\rangle =\frac{1-p^{\alpha }}{%
1-p^{-\alpha -n}}\int_{\mathbb{Q}_{p}^{n}}||x||_{p}^{-\alpha -n}(\varphi
(x)-\varphi (0))\,d^{n}x.  \label{7}
\end{equation}

\begin{lemma}[{\protect\cite[Chap. III, Theorem 4.5]{TA}}]
As elements of $S^{\prime }(\mathbb{Q}_{p}^{n})$, $\left( \mathcal{F}\mathit{%
k_{\alpha }}\right) \left( x\right) $ equals $||x||_{p}^{-\alpha }$, $\alpha
\neq n$.
\end{lemma}

\begin{definition}
The Taibleson pseudo-differential operator $D_{T}^{\alpha }$, $\alpha >0$,
is defined as 
\begin{equation*}
(D_{T}^{\alpha }\varphi )(x)=\mathcal{F}_{\xi \rightarrow x}^{-1}\left(
||\xi ||_{p}^{\alpha }\mathcal{F}_{x\rightarrow \xi }\varphi \right) \text{,
for }\varphi \in S(\mathbb{Q}_{p}^{n})\text{.}
\end{equation*}
\end{definition}

As a consequence of the previous lemma and (\ref{7}), we have 
\begin{align}
\left( D_{T}^{\alpha }\varphi \right) \left( x\right) & =\left( \mathit{k}%
_{-\alpha }\ast \varphi \right) \left( x\right) =  \notag \\
& \frac{1-p^{\alpha }}{1-p^{-\alpha -n}}\int_{\mathbb{Q}%
_{p}^{n}}||y||_{p}^{-\alpha -n}(\varphi (x-y)-\varphi (x))\,d^{n}y.
\label{8}
\end{align}%
The right-hand side of (\ref{8}) makes sense for a wider class of functions,
for example, for locally constant functions $\varphi (x)$ satisfying 
\begin{equation*}
\int_{||x||_{p}\geq 1}||x||_{p}^{-\alpha -n}|\varphi (x)|\,d^{n}x<\infty .
\end{equation*}

\section{The $p$-adic Heat Equation and the Taibleson Operator}

In this paper we consider the following Cauchy problem: 
\begin{equation}
\begin{cases}
\frac{\partial u(x,t)}{\partial t}+a(D_{T}^{\alpha }u)(x,t)=f(x,t),\quad
x\in \mathbb{Q}_{p}^{n},\quad t\in (0,T], \\ 
\\ 
u(x,0)=\varphi (x),%
\end{cases}
\label{Heat}
\end{equation}%
where $a>0$, $\alpha >0$ and $D_{T}^{\alpha }$ is the Taibleson operator. In
this section we show that (\ref{Heat}) is a multi-dimensional analog of the $%
p$-adic heat equation introduced in \cite{VVZ}.

\subsection{The Fundamental Solution}

The\textit{\ fundamental solution} for the Cauchy problem (\ref{Heat}) is
defined as 
\begin{equation}
Z(x,t):=\int_{\mathbb{Q}_{p}^{n}}\Psi (x\cdot \xi )e^{-at||\xi
||_{p}^{\alpha }}\,d^{n}\xi .  \label{fundamental solution}
\end{equation}

\begin{lemma}
\label{lemma1}The fundamental solution has the following properties:

\noindent 1) $Z(x,t)=(1-p^{-n})||x||_{p}^{-n}\sum_{k=0}^{\infty
}q^{-kn}e^{-at(q^{-k}||x||_{p}^{-1})^{\alpha
}}-||x||_{p}^{-n}e^{-at(p||x||_{p}^{-1})^{\alpha }};$

\noindent 2) $Z(x,t)=\sum_{m=1}^{\infty }\frac{(-1)^{m}}{m!}\,\frac{%
1-p^{\alpha m}}{1-p^{-\alpha m-n}}\,(at)^{m}||x||_{p}^{-\alpha m-n}$ for $%
x\neq 0;$

\noindent 3) $Z(x,t)\geq 0$, for all $x\in \mathbb{Q}_{p}^{n}$,$\quad t\in
(0,T].$
\end{lemma}

\begin{proof}
1) By expanding $Z(x,t)$ as 
\begin{equation*}
Z(x,t)=\sum_{k=-\infty }^{\infty }\int_{||\xi ||_{p}=p^{k}}\Psi (x\cdot \xi
)e^{-at||\xi ||_{p}^{\alpha }}\,d^{n}\xi ,
\end{equation*}

and applying 
\begin{equation*}
\int_{||\xi ||_{p}=p^{k}}\Psi (x\cdot \xi )\,d^{n}\xi =\left\{ 
\begin{array}{lll}
p^{kn}(1-p^{-n}), & \text{if} & ||x||_{p}\leq p^{-k} \\ 
&  &  \\ 
-p^{kn}p^{-n}, & \text{if} & \left\Vert x\right\Vert _{p}=p^{-k+1} \\ 
&  &  \\ 
0, & \text{if} & ||x||_{p}>p^{-k+1},%
\end{array}%
\right.
\end{equation*}%
(c.f. Lemma 4.1 in \cite[Chap. III]{TA}), we obtain 
\begin{equation}
Z(x,t)=(1-p^{-n})||x||_{p}^{-n}\sum_{k=0}^{\infty
}p^{-kn}e^{-at(p^{-k}||x||_{p}^{-1})^{\alpha
}}-||x||_{p}^{-n}e^{-at(p||x||_{p}^{-1})^{\alpha }}.  \label{9}
\end{equation}%
Note that by the previous expansion $Z(x,t)$ is a real-valued function.

2) By using the Taylor expansion of $e^{x}$ in (\ref{9}), and exchanging the
order of summation, and sum the geometric progression, we find that 
\begin{equation*}
Z(x,t)=\sum_{m=1}^{\infty}\frac{(-1)^{m}}{m!}\,\frac{1-p^{\alpha m}}{%
1-p^{-\alpha m-n}}\,(at)^{m}||x||_{p}^{-\alpha m-n}\text{, for }x\neq0.
\end{equation*}

3) Let $\Omega _{l}\left( x\right) $ denote the characteristic function of
the ball $B_{-l}^{n}\left( 0\right) $. Then $\mathcal{F}\Omega
_{l}=p^{-nl}\Omega _{-l}$. The last part follows from this observation by
means of the following calculation: 
\begin{align*}
Z(x,t)& =\sum_{l=-\infty }^{\infty }e^{-atp^{l\alpha }}\int_{||\xi
||_{p}=p^{l}}\Psi (x\cdot \xi )\,d^{n}\xi \\
& =\sum_{l=-\infty }^{\infty }e^{-atp^{l\alpha }}(p^{n\left( l\right)
}\Omega _{-l}(x)-p^{n\left( l-1\right) }\Omega _{-l+1}(x)) \\
& =\sum_{l=-\infty }^{\infty }p^{nl}(e^{-atp^{l\alpha }}-e^{-atp^{\left(
l+1\right) \alpha }})\Omega _{-l}(x)\geq 0
\end{align*}
\end{proof}

\begin{lemma}
\label{lemma2}%
\begin{equation}
Z(x,t)\leq Ct(t^{1/\alpha }+||x||_{p})^{-\alpha -n}\text{, }t>0\text{, }x\in 
\mathbb{Q}_{p}^{n}\text{.}  \label{10}
\end{equation}
\end{lemma}

\begin{proof}
Let $l$ an integer such that $p^{l-1}\leq t^{1/\alpha}\leq p^{l}$. Then 
\begin{align}
Z(x,t) & \leq\int_{\mathbb{Q}_{p}^{n}}e^{-at||\xi||_{p}^{\alpha}}\,d^{n}\xi%
\leq\int_{\mathbb{Q}_{p}^{n}}e^{-ap^{\alpha(l-1)}||\xi||_{p}^{\alpha}}%
\,d^{n}\xi=\int_{\mathbb{Q}_{p}^{n}}e^{-a||p^{-(l-1)}\xi||_{p}^{\alpha}}%
\,d^{n}\xi  \notag \\
& =p^{-(l-1)n}\int_{\mathbb{Q}_{p}^{n}}e^{-a||\eta||_{p}^{\alpha}}\,d%
\eta=C_{0}\left( \alpha\right) p^{-n}p^{-ln}\leq C_{1}t^{-n/\alpha }.
\label{11}
\end{align}
On the other hand, if $||x||_{p}\geq t^{1/\alpha}$, by applying Lemma \ref%
{lemma1} (2), we have 
\begin{equation}
Z(x,t)\leq||x||_{p}^{-n}\sum_{m=1}^{\infty}\frac{C_{2}^{m}}{m!}%
\,(t||x||_{p}^{-\alpha})^{m}\leq C_{3}t||x||_{p}^{-\alpha-n}.  \label{12}
\end{equation}
The result follows from (\ref{11}-\ref{12})\ as follows. If $||x||_{p}\geq
t^{1/\alpha}$, by (\ref{12}), 
\begin{equation*}
Z(x,t)\leq C_{3}t||x||_{p}^{-\alpha-n}\leq2^{\alpha+n}C_{3}t(t^{1/\alpha
}+||x||_{p})^{-\alpha-n}.
\end{equation*}
If $||x||_{p}<t^{1/\alpha}$, by (\ref{11}), 
\begin{equation*}
Z(x,t)\leq
C_{1}t^{-n/\alpha}\leq2^{\alpha+n}C_{1}t(t^{1/\alpha}+||x||_{p})^{-\alpha-n}.
\end{equation*}
\end{proof}

Inequality (\ref{10}) shows in particular that the function $Z\left(
x,t\right) $ belongs, with respect to $x$, to $L_{1}(\mathbb{Q}_{p}^{n})\cap
L_{2}(\mathbb{Q}_{p}^{n})$.

\begin{corollary}
\label{cor1}%
\begin{equation}
\dint\limits_{\mathbb{Q}_{p}^{n}}Z(x,t)d^{n}x=1.  \label{14}
\end{equation}
\end{corollary}

\subsection{The Spaces $\mathfrak{M}_{\protect\lambda }$ and
Pseudo-differentiability of the Fundamental Solution}

\begin{definition}
Denote by $\mathfrak{M}_{\lambda }$, $\lambda >0$, the set of the
complex-valued locally constant functions $\varphi (x)$ on $\mathbb{Q}%
_{p}^{n}$ such that 
\begin{equation*}
|\varphi (x)|\leq C\left( \varphi \right) (1+||x||_{p}^{\lambda }).
\end{equation*}%
If the function $\varphi $ depends also on a parameter $t$, we shall say
that $\varphi \in \mathfrak{M}_{\lambda }$ \textit{uniformly with respect to 
}$t$, if its constant $C$ and its exponent of local constancy $l\left(
\varphi \right) $ do not depend on $t$.
\end{definition}

\begin{lemma}
\label{lemma3}If $\varphi \in \mathfrak{M}_{\lambda }$, $\lambda <\alpha $,
with $\alpha $\ as in (\ref{Heat}), then 
\begin{equation}
\lim_{t\rightarrow 0^{+}}\int_{\mathbb{Q}_{p}^{n}}Z(x-\xi ,t)\varphi (\xi
)\,d^{n}\xi =\varphi (x).  \label{15}
\end{equation}
\end{lemma}

\begin{proof}
By Corollary (\ref{cor1}) and Lemmas \ref{lemma1} (part 3) and \ref{lemma2}
we have 
\begin{equation*}
\left\vert \int_{\mathbb{Q}_{p}^{n}}Z(x-\xi ,t)\varphi (\xi )\,d^{n}\xi
-\varphi (x)\right\vert =\left\vert \int_{\mathbb{Q}_{p}^{n}}Z(x-\xi
,t)\left( \varphi (\xi )-\varphi (x)\right) \,d^{n}\xi \right\vert
\end{equation*}%
\begin{align*}
& \leq \int_{\mathbb{Q}_{p}^{n}}Z(x-\xi ,t)|\varphi (\xi )-\varphi
(x)|\,d^{n}\xi \\
& \leq C\int_{\mathbb{Q}_{p}^{n}}t(t^{1/\alpha }+||x-\xi ||_{p})^{-\alpha
-n}|\varphi (\xi )-\varphi (x)|\,d^{n}\xi :=I\left( x,t\right)
\end{align*}%
Let $\eta $ be the exponent of locally constancy of $\varphi $. Since $%
\varphi \in \mathfrak{M}_{\lambda }$, $\lambda <\alpha $, we can re-write $%
I\left( x,t\right) $ as follows:%
\begin{align*}
I\left( x,t\right) & =C\int_{||\xi -x||_{p}>p^{\eta }}t(t^{1/\alpha }+||\xi
-x||_{p})^{-\alpha -n}|\varphi (\xi )-\varphi (x)|\,d^{n}\xi \\
& \\
& \leq I_{1}(x,t)+I_{2}(x,t),
\end{align*}%
with 
\begin{align*}
& I_{1}\left( x,t\right) :=C_{1}t\int\limits_{||\xi -x||_{p}>p^{\eta }}\frac{%
1+||\xi ||_{p}^{\lambda }}{(t^{1/\alpha }+||x-\xi ||_{p})^{\alpha +n}}%
\,d^{n}\xi , \\
& I_{2}(x,t):=Ct|\varphi (x)|\int\limits_{||\xi -x||_{p}>p^{\eta
}}(t^{1/\alpha }+||\xi -x||_{p})^{-\alpha -n}\,d^{n}\xi .
\end{align*}%
Now, since $\alpha >0$, and $t>0$, 
\begin{equation*}
I_{2}(x,t)\leq C_{2}t|\varphi (x)|,
\end{equation*}%
and since $\lambda <\alpha $, 
\begin{equation*}
I_{1}\left( x,t\right) \leq C_{1}t\left( C_{3}+\int_{||\tau ||_{p}>p^{\eta }}%
\frac{||x-\tau ||_{p}^{\lambda }}{||\tau ||_{p}{}^{\alpha +n}}\,d^{n}\xi
\right) \leq
\end{equation*}%
\begin{equation*}
C_{1}t\left( C_{3}+\int_{p^{\eta }<||\tau ||_{p}^{\eta }\leq \left\Vert
x\right\Vert _{p}}\frac{||x-\tau ||_{p}^{\lambda }}{||\tau ||_{p}{}^{\alpha
+n}}\,d^{n}\xi +\int_{||\tau ||_{p}>\left\Vert x\right\Vert _{p}}\frac{%
||x-\tau ||_{p}^{\lambda }}{||\tau ||_{p}{}^{\alpha +n}}\,d^{n}\xi \right) =
\end{equation*}%
\begin{equation*}
C_{1}t\left( C_{4}\left( x\right) +\int_{||\tau ||_{p}>\left\Vert
x\right\Vert _{p}}\frac{1}{||\tau ||_{p}{}^{\alpha -\lambda +n}}\,d^{n}\xi
\right) =C_{5}\left( x\right) t.
\end{equation*}%
Therefore 
\begin{equation*}
\lim_{t\rightarrow 0^{+}}\left\vert \int_{\mathbb{Q}_{p}^{n}}Z(x-\xi
,t)\varphi (\xi )\,d^{n}\xi -\varphi (x)\right\vert \leq \lim_{t\rightarrow
0^{+}}C_{6}\left( x\right) t=0.
\end{equation*}
\end{proof}

For further reference we summarize the properties of the fundamental
solution in the following proposition.

\begin{proposition}
\label{prop1A}The fundamental solution has the following properties:

\noindent (1) $Z(x,t)\geq 0$, for all $x\in \mathbb{Q}_{p}^{n}$,$\quad t\in
(0,T].$

\noindent (2) $\tint\nolimits_{\mathbb{Q}_{p}^{n}}Z\left( x,t\right) $ $%
d^{n}x=1$, for any $t>0$;

\noindent (3) if $\varphi \in S\left( \mathbb{Q}_{p}^{n}\right) $, then $%
\lim_{\left( x,t\right) \rightarrow \left( x_{0},0\right) }\tint\nolimits_{%
\mathbb{Q}_{p}^{n}}Z\left( x-\eta ,t\right) \varphi \left( \eta \right)
d^{n}\eta =\varphi \left( x_{0}\right) $;

\noindent (4) $Z\left( x,t+t^{\prime }\right) =\tint\nolimits_{\mathbb{Q}%
_{p}^{n}}Z\left( x-y,t\right) Z\left( y,t^{\prime }\right) d^{n}y$, for $t$, 
$t^{\prime }>0$.
\end{proposition}

\begin{proof}
(1), (2), and (3) are already established (c.f. Lemma \ref{lemma1}-part (3),
Corollary \ref{cor1}, and Lemma \ref{lemma3}). The last assertion is proved
as follows: since $e^{-at||\xi ||_{p}^{\alpha }}\in L^{1}\left( \mathbb{Q}%
_{p}^{n}\right) $, 
\begin{align*}
\int_{\mathbb{Q}_{p}^{n}}Z(x-y,t_{1})Z(y,t_{2})\,d^{n}y& =\mathcal{F}%
^{-1}\left( \mathcal{F}(Z(y,t_{1})\ast Z(y,t_{2}))\right) \\
& =\mathcal{F}^{-1}\left( e^{-at_{1}||\xi ||_{p}^{\alpha }}e^{-at_{2}||\xi
||_{p}^{\alpha }}\right) \\
& =Z(x,t_{1}+t_{2}).
\end{align*}
\end{proof}

\begin{proposition}
\label{lemma 4} If $b>0$, $0\leq \lambda <\alpha $, and $x\in \mathbb{Q}%
_{p}^{n}$, then 
\begin{equation*}
I(b,x)=\int_{\mathbb{Q}_{p}^{n}}\left( b+||x-\xi ||_{p}\right) ^{-\alpha
-n}||\xi ||_{p}^{\lambda }\,d^{n}\xi \leq Cb^{-\alpha }\left(
1+||x||_{p}^{\lambda }\right) ,
\end{equation*}%
where the constant $C$ does not depend on $b$, $x$.
\end{proposition}

\begin{proof}
Let $m$ be an integer such that $p^{m-1}\leq b\leq p^{m}$. Then 
\begin{equation*}
\left( b+||x-\xi ||_{p}\right) ^{-\alpha -n}\leq \left( p^{m-1}+||x-\xi
||_{p}\right) ^{-\alpha -n},
\end{equation*}%
and%
\begin{equation*}
I(b,x)\leq I(p^{m-1},x)=\int_{\mathbb{Q}_{p}^{n}}\left( p^{m-1}|_{p}+||x-\xi
||_{p}\right) ^{-\alpha -n}||\xi ||_{p}^{\lambda }\,d^{n}\xi
\end{equation*}%
\begin{align}
& =p^{(m-1)(-\alpha -n)}\int_{\mathbb{Q}_{p}^{n}}\left(
1+||p^{m-1}x-p^{m-1}\xi ||_{p}\right) ^{-\alpha -n}||\xi ||_{p}^{\lambda
}\,d^{n}\xi  \notag \\
& =p^{(m-1)(\lambda -\alpha )}\int_{\mathbb{Q}_{p}^{n}}\left(
1+||p^{m-1}x-\eta ||_{p}\right) ^{-\alpha -n}||\eta ||_{p}^{\lambda
}\,d^{n}\eta  \notag \\
& =p^{(m-1)(\lambda -\alpha )}I(1,p^{m-1}x).  \label{17}
\end{align}

Let $p^{m-1}x=y$, $||y||_{p}=p^{l}$. We have 
\begin{equation*}
I(1,y)=I_{1}(y)+I_{2}(y)+I_{3}(y),
\end{equation*}%
where 
\begin{align*}
I_{1}(y)& =\sum_{k=-\infty }^{l-1}\int_{||\eta ||_{p}=p^{k}}\left(
1+||y-\eta ||_{p}\right) ^{-\alpha -n}||\eta ||_{p}^{\lambda }\,d^{n}\eta ,
\\
I_{2}(y)& =\int_{||\eta ||_{p}=p^{l}}\left( 1+||y-\eta ||_{p}\right)
^{-\alpha -n}||\eta ||_{p}^{\lambda }\,d^{n}\eta , \\
I_{3}(y)& =\sum_{k=l+1}^{\infty }\int_{||\eta ||_{p}=p^{k}}\left( 1+||y-\eta
||_{p}\right) ^{-\alpha -n}||\eta ||_{p}^{\lambda }\,d^{n}\eta .
\end{align*}

The results follows from the following estimations:

\textbf{Claim A.} $I_{1}(y)\leq C_{0}(1+||y||_{p})^{-\alpha
-n}||y||_{p}^{\lambda +n};$

\textbf{Claim B. }$I_{2}(y)\leq C_{1}||y||_{p}^{\lambda }$;

\textbf{Claim C.} $I_{2}(y)\leq C_{2}$.

Indeed, from the claims we have $I(1,y)\leq C_{3}(1+||y||_{p}^{\lambda })$,
and by (\ref{17}), 
\begin{align*}
I(b,x)& \leq C_{3}p^{(m-1)(\lambda -\alpha )}(1+p^{(1-m)\lambda
}||x||_{p}^{\lambda }) \\
& \leq C_{3}p^{-m\alpha }(1+||x||_{p}^{\lambda })\leq Cb^{-\alpha }\left(
1+||x||_{p}^{\lambda }\right) .
\end{align*}

We now prove the announced claims.

\textbf{Proof of Claim A.} 
\begin{align*}
I_{1}(y)& =\sum_{k=-\infty }^{l-1}\int_{||\eta ||_{p}=p^{k}}\left(
1+||y-\eta ||_{p}\right) ^{-\alpha -n}||\eta ||_{p}^{\lambda }\,d^{n}\eta ,
\\
& =(1-p^{-n})(1+||y||_{p})^{-\alpha -n}\sum_{k=-\infty }^{l-1}p^{(\lambda
+n)k} \\
& \leq C_{0}(1+||y||_{p})^{-\alpha -n}||y||_{p}^{\lambda +n},
\end{align*}%
where 
\begin{equation*}
C_{0}=\frac{(1-p^{-n})p^{-\lambda -n}}{1-p^{-\lambda -n}}.
\end{equation*}

\textbf{Proof of Claim B. }Let $\widetilde{y}\in \mathbb{Q}_{p}$ such that $|%
\widetilde{y}|_{p}=p^{l}=||y||_{p}$, then 
\begin{align*}
I_{2}(y)& =\int_{||\eta ||_{p}=p^{l}}\left( 1+||y-\eta ||_{p}\right)
^{-\alpha -n}||\eta ||_{p}^{\lambda }\,d^{n}\eta \\
& =||y||_{p}^{\lambda }\int_{||\eta ||_{p}=p^{l}}\left( 1+|\widetilde{y}%
|_{p}||\widetilde{y}^{-1}y-\widetilde{y}^{-1}\eta ||_{p}\right) ^{-\alpha
-n}\,d^{n}\eta \\
& =||y||_{p}^{\lambda -\alpha }\int_{||\eta ||_{p}=1}\left(
||y||_{p}^{-1}+||u-\eta ||_{p}\right) ^{-\alpha -n}\,d^{n}\eta \text{, with
\ }u=\widetilde{y}^{-1}y.
\end{align*}%
We set 
\begin{equation*}
A_{m}=\{\eta \in \mathbb{Q}_{p}^{n}\mid ||\eta ||_{p}=1\text{ and }||u-\eta
||_{p}=p^{-m}\},\text{ for }m\in \mathbb{N},
\end{equation*}%
and for $I$ non-empty subset of $\{1,2,\dotsc ,n\}$, 
\begin{equation*}
A_{m,I}=\{\eta \in A_{m}\mid |u_{i}-\eta _{i}|_{p}=p^{-m}\text{ for }i\in I%
\text{ and }|u_{i}-\eta _{i}|_{p}<p^{-m}\text{ for }i\notin I\},
\end{equation*}%
where \ $u=\left( u_{1},\ldots ,u_{n}\right) $, $\eta =\left( \eta
_{1},\ldots ,\eta _{n}\right) \in \mathbb{Q}_{p}^{n}$, with $||\eta
||_{p}=||u||_{p}=1$.

With this notation we have $A_{m}\subseteq \bigcup_{I}A_{m,I}$, 
\begin{equation*}
vol(A_{m,I})\leq (p^{-m}(1-p^{-1}))^{|I|}(p^{-m-1})^{n-|I|},
\end{equation*}%
here $|I|$ denotes the cardinality of $I$, then 
\begin{equation*}
vol(A_{m})\leq \sum_{|I|=0}^{n}\dbinom{n}{|I|}%
(p^{-m}(1-p^{-1}))^{|I|}(p^{-m-1})^{n-|I|}=p^{-mn},
\end{equation*}%
and 
\begin{align*}
I_{2}(y)& =||y||_{p}^{\lambda -\alpha }\sum_{m=0}^{\infty
}\int_{A_{m}}\left( ||y||_{p}^{-1}+||u-\eta ||_{p}\right) ^{-\alpha
-n}\,d^{n}\eta \\
& \leq ||y||_{p}^{\lambda -\alpha }\sum_{m=0}^{\infty }\left(
||y||_{p}^{-1}+p^{-m}\right) ^{-\alpha -n}p^{-mn} \\
& =\frac{||y||_{p}^{\lambda -\alpha }}{1-p^{-n}}\sum_{m=0}^{\infty
}\int_{||\eta ||_{p}=p^{-m}}\left( ||y||_{p}^{-1}+||\eta ||_{p}\right)
^{-\alpha -n}\,d^{n}\eta \\
& =\frac{||y||_{p}^{\lambda -\alpha }}{1-p^{-n}}\int_{||\eta ||_{p}\leq
1}\left( ||y||_{p}^{-1}+||\eta ||_{p}\right) ^{-\alpha -n}\,d^{n}\eta \\
& \leq C_{1}^{\prime }||y||_{p}^{\lambda -\alpha }\int_{\mathbb{Q}%
_{p}^{n}}\left( ||y||_{p}^{-1}+||\eta ||_{p}\right) ^{-\alpha -n}\,d^{n}\eta
\\
& =C_{1}^{\prime }||y||_{p}^{\lambda -\alpha }\int_{\mathbb{Q}%
_{p}^{n}}\left( ||y||_{p}^{-1}+||y||_{p}^{-1}||\widetilde{y}\eta
||_{p}\right) ^{-\alpha -n}\,d^{n}\eta \\
& =C_{1}^{\prime }||y||_{p}^{\lambda +n}\int_{\mathbb{Q}_{p}^{n}}\left( 1+||%
\widetilde{y}\eta ||_{p}\right) ^{-\alpha -n}\,d^{n}\eta \\
& =C_{1}^{\prime }||y||_{p}^{\lambda }\int_{\mathbb{Q}_{p}^{n}}\left(
1+||\tau ||_{p}\right) ^{-\alpha -n}\,d\tau \leq C_{1}||y||_{p}^{\gamma }.
\end{align*}

\textbf{Proof of Claim C.} 
\begin{align*}
I_{3}(y) & =\sum_{k=l+1}^{\infty}\int_{||\eta||_{p}=p^{k}}\left(
1+||\eta||_{p}\right) ^{-\alpha-n}||\eta||_{p}^{\lambda}\,d^{n}\eta, \\
& \leq\int_{\mathbb{Q}_{p}^{n}}\left( 1+||\eta||_{p}\right) ^{-\alpha
-n}||\eta||_{p}^{\lambda}\,d^{n}\eta=C.
\end{align*}
\end{proof}

\begin{lemma}
\label{lemma6} If $\alpha >0$, then 
\begin{equation}
||x||_{p}^{\alpha }=\frac{1}{\Gamma _{p}^{\left( n\right) }(-\alpha )}\int_{%
\mathbb{Q}_{p}^{n}}||y||_{p}^{-\alpha -n}\left( \Psi (-x\cdot y)-1\right)
\,d^{n}y  \label{eq lemma6}
\end{equation}%
for all $x\in \mathbb{Q}_{p}^{n}$.
\end{lemma}

\begin{proof}
The proof is a slightly variation of the proof of Proposition 2.3 in \cite%
{Koch1}.
\end{proof}

\begin{lemma}
\label{lemma7} Let $0<\gamma \leq \alpha $, then 
\begin{equation*}
(D_{T}^{\gamma }Z)(x,t)=\int_{\mathbb{Q}_{p}^{n}}\Psi (x\cdot \eta )||\eta
||_{p}^{\gamma }e^{-at||\eta ||_{p}^{\alpha }}\,d^{n}\eta .
\end{equation*}
\end{lemma}

\begin{proof}
By Lemma \ref{lemma1} (2), $Z(x-y,t)=Z(x,t)$, for $||y||<||x||$. Then we can
use (\ref{8}) to calculate$(D_{T}^{\gamma }Z)(x,t)$:%
\begin{align*}
(D_{T}^{\gamma }Z)(x,t)& =\frac{1}{\Gamma _{p}^{\left( n\right) }(-\gamma )}%
\int_{\mathbb{Q}_{p}^{n}}||y||_{p}^{-\gamma -n}(Z(x-y,t)-Z(x,t))\,d^{n}y \\
& =\frac{1}{\Gamma _{p}^{\left( n\right) }(-\gamma )}\int_{||y||_{p}\geq
||x||_{p}}||y||_{p}^{-\gamma -n}(Z(x-y,t)-Z(x,t))\,d^{n}y.
\end{align*}%
We now use Lemma \ref{lemma2} to obtain%
\begin{eqnarray}
\left\vert (D_{T}^{\gamma }Z)(x,t)\right\vert &\leq &\left\vert \frac{1}{%
\Gamma _{p}^{\left( n\right) }(-\gamma )}\right\vert \int_{||y||_{p}\geq
||x||_{p}}\left( Ct||y||_{p}^{-\gamma -\alpha -2n}+Z(x,t)||y||_{p}^{-\gamma
-n}\right) \,d^{n}y  \notag \\
&<&\infty .  \label{20}
\end{eqnarray}%
This shows that $(D_{T}^{\gamma }Z)(x,t)$ exists. We now compute this
function explicitly.

We set 
\begin{equation*}
Z^{(m)}(x,t):=\int_{||\xi ||_{p}\leq p^{m}}\Psi (x\cdot \xi )e^{-at||\xi
||_{p}^{\alpha }}\,d^{n}\xi .
\end{equation*}%
Then $Z^{(m)}(x,t)$ is bounded and locally constant as function of $x$, the
exponent of local constancy is $m$. From these observations by using Lemma %
\ref{lemma1} (2), and (\ref{8}) we calculate $(D_{T}^{\gamma }Z)(x,t)$ as
follows:%
\begin{equation*}
(D_{T}^{\gamma }Z^{(m)})(x,t)=\frac{1}{\Gamma _{p}^{\left( n\right)
}(-\gamma )}\int_{\mathbb{Q}_{p}^{n}}||y||_{p}^{-\gamma
-n}(Z^{(m)}(x-y,t)-Z^{(m)}(x,t))\,d^{n}y
\end{equation*}%
\begin{equation*}
=\frac{1}{\Gamma _{p}^{\left( n\right) }(-\gamma )}%
\int_{||y||_{p}>p^{-m}}||y||_{p}^{-\gamma
-n}(Z^{(m)}(x-y,t)-Z^{(m)}(x,t))\,d^{n}y
\end{equation*}%
\begin{equation*}
=\frac{1}{\Gamma _{p}^{\left( n\right) }(-\gamma )}%
\int_{||y||_{p}>p^{-m}}||y||_{p}^{-\gamma -n}\int_{||\eta ||_{p}\leq
p^{m}}e^{-at||\eta ||_{p}^{\alpha }}\Psi (x\cdot \eta )\left( \Psi (-y\cdot
\eta )-1\right) \,d^{n}\eta \,d^{n}y
\end{equation*}%
\begin{eqnarray*}
&=&\int_{||\eta ||_{p}\leq p^{m}}e^{-at||\eta ||^{\alpha }}\Psi (x\cdot \eta
)\times \\
&&\left( \frac{1}{\Gamma _{p}^{\left( n\right) }(-\gamma )}%
\int_{||y||_{p}>p^{-m}}||y||_{p}^{-\gamma -n}(\Psi (-y\cdot \eta
)-1)d^{n}y\right) \,d^{n}\eta .
\end{eqnarray*}%
Note that if $||y||_{p}\leq p^{-m}$, then $\Psi (-y\cdot \eta )=1$ for all $%
\eta $ such that $||\eta ||_{p}\leq p^{m}$, using this observation and Lemma
(\ref{lemma6}), $(D_{T}^{\gamma }Z^{(m)})(x,t)$ becomes 
\begin{align*}
& \int_{||\eta ||_{p}\leq p^{m}}e^{-at||\eta ||_{p}^{\alpha }}\Psi (x\cdot
\eta )\left( \frac{1}{\Gamma _{p}^{\left( n\right) }(-\gamma )}\int_{\mathbb{%
Q}_{p}^{n}}||y||_{p}^{-\gamma -n}(\Psi (-y\cdot \eta )-1)\,d^{n}y\right)
\,d^{n}\eta \\
& =\int_{||\eta ||_{p}\leq p^{m}}e^{-at||\eta ||^{\alpha }}\Psi (x\cdot \eta
)||\eta ||_{p}^{\gamma }\,d^{n}\eta .
\end{align*}%
By the dominated convergence theorem and ( \ref{20}) we have 
\begin{equation*}
(D_{T}^{\gamma }Z)(x,t)=\int_{\mathbb{Q}_{p}^{n}}e^{-at||\eta ||_{p}^{\alpha
}}\Psi (x\cdot \eta )||\eta ||_{p}^{\gamma }\,d^{n}\eta .
\end{equation*}
\end{proof}

\begin{lemma}
\label{lemma8}%
\begin{equation*}
\frac{\partial Z}{\partial t}(x,t)=-a\int_{\mathbb{Q}_{p}^{n}}\Psi (x\cdot
\xi )||\xi ||_{p}^{\alpha }e^{-at||\xi ||_{p}^{\alpha }}\,d^{n}\xi ;
\end{equation*}%
\begin{equation*}
\frac{\partial Z}{\partial t}(x,t)=-a(D_{T}^{\gamma }Z)(x,t)\text{, for \ }%
0<\gamma \leq \alpha .
\end{equation*}
\end{lemma}

\begin{proof}
The first part follows by applying the dominated convergence theorem. The
second part follows from the first one by Lemma \ref{lemma7}.
\end{proof}

\begin{lemma}
\label{lemma9}%
\begin{equation*}
\left\vert \frac{\partial Z}{\partial t}(x,t)\right\vert \leq C\left(
t^{1/\alpha }+||x||_{p}\right) ^{-\alpha -n};
\end{equation*}%
\begin{equation*}
\left\vert (D_{T}^{\gamma }Z)(x,t)\right\vert \leq C\left( t^{1/\alpha
}+||x||_{p}\right) ^{-\gamma -n}.
\end{equation*}
\end{lemma}

\begin{proof}
The proof uses the same reasonings as the one given in the proof of Lemma %
\ref{lemma2}.
\end{proof}

\begin{corollary}
\label{cor2} 
\begin{equation*}
\int_{\mathbb{Q}_{p}^{n}}(D_{T}^{\gamma }Z)(x,t)\,d^{n}x=0
\end{equation*}
\end{corollary}

\subsection{The Cauchy Problem for the multidimensional $p$-adic Heat
Equation}

\begin{theorem}
\label{Thm 1} Let $\varphi (x),$ $f(x,t)\in \mathfrak{M}_{\lambda }$, $0\leq
\lambda <\alpha $ be continuous functions. Then the Cauchy problem 
\begin{equation}
\begin{cases}
\frac{\partial u(x,t)}{\partial t}+a(D_{T}^{\alpha }u)(x,t)=f(x,t),\quad
x\in \mathbb{Q}_{p}^{n},\quad t\in (0,T], \\ 
\\ 
u(x,0)=\varphi (x),%
\end{cases}
\label{Cauchy}
\end{equation}%
with $a>0$, $\alpha >0$, has a continuous solution in $\mathfrak{M}_{\lambda
}$ given by 
\begin{equation*}
u(x,t)=\int_{\mathbb{Q}_{p}^{n}}Z(x-\xi ,t)\varphi (\xi )\,d^{n}\xi
+\int_{0}^{t}\left( \int_{\mathbb{Q}_{p}^{n}}Z(x-\xi ,t-\tau )f(\xi ,\tau
)\,d^{n}\xi \right) \,d\tau .
\end{equation*}
\end{theorem}

The proof of the theorem will be accomplished through the following lemmas.

We set 
\begin{equation*}
u_{1}(x,t):=\int_{\mathbb{Q}_{p}^{n}}Z(x-\xi ,t)\varphi (\xi )\,d^{n}\xi ,%
\text{ \ and }
\end{equation*}%
\begin{equation*}
u_{2}(x,t):=\int_{0}^{t}\left( \int_{\mathbb{Q}_{p}^{n}}Z(x-\xi ,t-\tau
)f(\xi ,\tau )\,d^{n}\xi \right) \,d\tau .
\end{equation*}

\begin{lemma}
\label{lemma10} $u(x,t)\in \mathfrak{M}_{\lambda }$ uniformly with respect
to $t$, and $u(x,t)$ satisfies the initial conditions of Theorem \ref{Thm 1}.
\end{lemma}

\begin{proof}
We first show that $u_{1}(x,t)\in \mathfrak{M}_{\lambda }$ uniformly with
respect to $t$. Since $\varphi $ is locally constant, there exist $l\in 
\mathbb{N}$ such that $\ \varphi (\xi +y)=\varphi (\xi )$ for any $%
\left\Vert y\right\Vert _{p}\leq p^{-l}$. By changing variables $y-\xi
=-\eta $ in $u_{1}(x,t)$ we that $u_{1}(x,t)$ is locally constant. Now using
Lemma \ref{lemma2} and Proposition \ref{lemma 4} we have $\left\vert
u_{1}(x,t)\right\vert \leq C\left( 1+\left\Vert x\right\Vert \right)
^{\lambda }$, and thus $u_{1}(x,t)\in \mathfrak{M}_{\lambda }$ uniformly
with respect to $t$.

By a similar reasoning \ one shows that $u_{2}(x,t)$ is locally constant in $%
x$, and that $\left\vert u_{2}(x,t)\right\vert \leq CT\left( 1+\left\Vert
x\right\Vert \right) ^{\lambda }$. Therefore $u(x,t)=u_{1}(x,t)+u_{2}(x,t)%
\in \mathfrak{M}_{\lambda }$ uniformly with respect to $t$.

We now show that $\lim_{t\rightarrow 0^{+}}$ $u(x,t)$ $=\varphi (x)$: by
Lemma \ref{lemma3}, $\lim_{t\rightarrow 0^{+}}$ $u_{1}(x,t)$ $=\varphi (x)$,
and $\lim_{t\rightarrow 0^{+}}$ $u_{2}(x,t)$ $=0$, since $\left\vert
u_{2}(x,t)\right\vert \leq Ct\left( 1+\left\Vert x\right\Vert \right)
^{\lambda }$, $t\leq T$.
\end{proof}

We now compute the partial derivatives of $u_{1}(x,t),$ $u_{2}(x,t)$ with
respect to $t$.

\begin{lemma}
\label{lemma11} 
\begin{equation*}
\dfrac{\partial u_{1}}{\partial t}(x,t)=\int_{\mathbb{Q}_{p}^{n}}\dfrac{%
\partial Z}{\partial t}(x-\xi ,t)\varphi (\xi )\,d^{n}\xi .
\end{equation*}
\end{lemma}

\begin{proof}
The results follows by applying the dominated convergence theorem.
\end{proof}

\begin{lemma}
\label{lemma12} 
\begin{equation*}
\dfrac{\partial u_{2}}{\partial t}(x,t)=\int_{0}^{t}\left( \int_{\mathbb{Q}%
_{p}^{n}}\dfrac{\partial Z}{\partial t}(x-\xi ,t-\tau )\bigl(f(\xi ,\tau
)-f(x,\tau )\bigr)\,d^{n}\xi \right) \,d\tau +f(x,t).
\end{equation*}
\end{lemma}

\begin{proof}
Let 
\begin{equation*}
u_{2,h}(x,t):=\int_{0}^{t-h}\left( \int_{\mathbb{Q}_{p}^{n}}Z(x-\xi ,t-\tau
)f(\xi ,\tau )\,d^{n}\xi \right) \,d\tau ,
\end{equation*}%
where $h$ is a small positive number. Then $\dfrac{u_{h}(x,t+t^{\prime
})-u_{h}(x,t)}{t^{\prime }}$ equals%
\begin{eqnarray}
&&\int_{0}^{t-h}\left( \int_{\mathbb{Q}_{p}^{n}}\dfrac{Z(x-\xi ,t+t^{\prime
}-\tau )-Z(x-\xi ,t-\tau )}{t^{\prime }}\,f(\xi ,\tau )\,d^{n}\xi \right)
d\tau  \notag \\
&&+\int_{t-h}^{t-h+t^{\prime }}\left( \int_{\mathbb{Q}_{p}^{n}}\dfrac{%
Z(x-\xi ,t+t^{\prime }-\tau )-Z(x-\xi ,t-\tau )}{t^{\prime }}\,f(\xi ,\tau
)\,d^{n}\xi \right) \,d\tau  \notag \\
&&+\frac{1}{t^{\prime }}\int_{t-h}^{t-h+t^{\prime }}\left( \int_{\mathbb{Q}%
_{p}^{n}}Z(x-\xi ,t-\tau )f(\xi ,\tau )\,d^{n}\xi \right) \,d\tau .
\label{25}
\end{eqnarray}

By taking $t^{\prime }\rightarrow 0^{+}$ the first integral in (\ref{25})
tends to 
\begin{equation*}
\int_{0}^{t-h}\left( \int_{\mathbb{Q}_{p}^{n}}\dfrac{\partial Z}{\partial t}%
(x-\xi ,t-\tau )f(\xi ,\tau )\,d^{n}\xi \right) \,d\tau ,
\end{equation*}%
and by using the continuity of the functions 
\begin{align*}
& \int_{\mathbb{Q}_{p}^{n}}\left( Z(x-\xi ,t+t^{\prime }-\tau )-Z(x-\xi
,t-\tau )\right) f(\xi ,\tau )\,d^{n}\xi ,\text{ }\ \text{and } \\
& \int_{\mathbb{Q}_{p}^{n}}Z(x-\xi ,t-\tau )f(\xi ,\tau )\,d^{n}\xi ,
\end{align*}%
with respect to $\tau $, the second integral in (\ref{25}) tends to zero,
and the third integral tends to%
\begin{equation*}
\int_{\mathbb{Q}_{p}^{n}}Z(x-\xi ,h)f(\xi ,t-h)\,d^{n}\xi .
\end{equation*}%
Hence 
\begin{align*}
\dfrac{\partial u_{2,h}}{\partial t}(x,t)& =\int_{0}^{t-h}\left( \int_{%
\mathbb{Q}_{p}^{n}}\dfrac{\partial Z}{\partial t}(x-\xi ,t-\tau )f(\xi ,\tau
)\,d^{n}\xi \right) \,d\tau \\
& +\int_{\mathbb{Q}_{p}^{n}}Z(x-\xi ,h)f(\xi ,t-h)\,d^{n}\xi .
\end{align*}%
This expression can be re-written as 
\begin{eqnarray}
\dfrac{\partial u_{2,h}}{\partial t}(x,t) &=&\int_{0}^{t-h}\left( \int_{%
\mathbb{Q}_{p}^{n}}\dfrac{\partial Z}{\partial t}(x-\xi ,t-\tau )(f(\xi
,\tau )-f(x,\tau ))\,d^{n}\xi \right) d\tau  \notag \\
&&+\int_{0}^{t-h}\left( \int_{\mathbb{Q}_{p}^{n}}\dfrac{\partial Z}{\partial
t}(x-\xi ,t-\tau )f(x,\tau )\,d^{n}\xi \right) d\tau  \notag \\
&&+\int_{\mathbb{Q}_{p}^{n}}Z(x-\xi ,h)(f(\xi ,t-h)-f(\xi ,t))\,d^{n}\xi 
\notag \\
&&+\int_{\mathbb{Q}_{p}^{n}}Z(x-\xi ,h)f(\xi ,t)\,d^{n}\xi .  \label{26}
\end{eqnarray}

The first integral contains no singularity at $t=\tau $ due to Lemma \ref%
{lemma9} and the local constancy of $f$. By Corollary \ref{cor2}, the second
integral in (\ref{26}) is equal to zero. The third integral can be written
as the sum of the integrals over $\left\{ \xi \in \mathbb{Q}_{p}^{n}\mid
\left\Vert \xi \right\Vert _{p}\leq p^{m}\right\} $ and its complement; one
integral is estimated on the basis of uniform continuity of $f$, \ while the
other contains no singularity. Hence this integral tends to zero as $h$
approaches zero from the right. By Lemma \ref{lemma3}, the fourth integral
tends to $f(x,t)$\ as $h\rightarrow 0^{+}$, therefore 
\begin{equation*}
\dfrac{\partial u_{2}}{\partial t}(x,t)=\int_{0}^{t}\left( \int_{\mathbb{Q}%
_{p}^{n}}\dfrac{\partial Z}{\partial t}(x-\xi ,t-\tau )(f(\xi ,\tau
)-f(x,t))\,d^{n}\xi \right) \,d\tau +f(x,t).
\end{equation*}
\end{proof}

As a consequence of Lemmas \ref{lemma11}-\ref{lemma12}\ we obtain:

\begin{proposition}
\label{prop2} 
\begin{eqnarray*}
\dfrac{\partial u}{\partial t}(x,t) &=&\int_{\mathbb{Q}_{p}^{n}}\dfrac{%
\partial Z}{\partial t}(x-\xi ,t)\varphi (\xi )\,d^{n}\xi \\
&&+\int_{0}^{t}\left( \int_{\mathbb{Q}_{p}^{n}}\dfrac{\partial Z}{\partial t}%
(x-\xi ,t-\tau )(f(\xi ,\tau )-f(x,t))\,d^{n}\xi \right) \,d\tau +f(x,t).
\end{eqnarray*}
\end{proposition}

We now consider the action of the operator $D_{T}^{\gamma }$, $0<\gamma \leq
\alpha $ upon $u(x,t)$. We first note that $(D_{T}^{\gamma }u)(x,t)$ is
defined if $\gamma >\lambda $. This follows from (\ref{8}) using $u(x,t)\in 
\mathfrak{M}_{\lambda }$.

\begin{lemma}
\label{claim 3} Let $\lambda <\gamma \leq \alpha $, then 
\begin{equation*}
(D_{T}^{\gamma }u_{1})(x,t)=\int_{\mathbb{Q}_{p}^{n}}(D_{T}^{\gamma
}Z)(x-\xi ,t)\varphi (\xi )\,d^{n}\xi .
\end{equation*}
\end{lemma}

\begin{proof}
Let $Z_{\gamma }(x,t):=(D_{T}^{\gamma }Z)(x,t)$ and 
\begin{equation}
Z_{\gamma ,l}(x,t):=\frac{1}{\Gamma _{p}^{\left( n\right) }(-\gamma )}%
\int_{||y||_{p}>p^{-l}}||y||_{p}^{-\gamma -n}\bigl(Z(x-y,t)-Z(x,t)\bigr)%
\,d^{n}y.  \label{26A}
\end{equation}%
By the Fubini \ theorem 
\begin{equation*}
\frac{1}{\Gamma _{p}^{\left( n\right) }(-\gamma )}%
\int_{||y||_{p}>p^{-l}}||y||_{p}^{-\gamma -n}\bigl(u_{1}(x-y,t)-u_{1}(x,t)%
\bigr)\,d^{n}y
\end{equation*}%
\begin{align*}
& =\frac{1}{\Gamma _{p}^{\left( n\right) }(-\gamma )}%
\int_{||y||_{p}>p^{-l}}||y||_{p}^{-\gamma -n}\left( \int_{\mathbb{Q}_{p}^{n}}%
\bigl(Z(x-y-\xi ,t)-Z(x-\xi ,t)\bigr)\varphi (\xi )\,d^{n}\xi \right)
\,d^{n}y \\
& =\int_{\mathbb{Q}_{p}^{n}}\varphi (\xi )\left( \frac{1}{\Gamma
_{p}^{\left( n\right) }(-\gamma )}\int_{||y||_{p}>p^{-l}}||y||_{p}^{-\gamma
-n}\bigl(Z(x-y-\xi ,t)-Z(x-\xi ,t)\bigr)\,d^{n}y\right) \,d^{n}\xi \\
& =\int_{\mathbb{Q}_{p}^{n}}Z_{\gamma ,l}(x-\xi ,t)\varphi (\xi )\,d^{n}\xi .
\end{align*}%
Let $m$ a fixed positive integer, then the last integral can expressed as 
\begin{equation*}
\int_{||x-\xi ||_{p}\geq p^{-m}}Z_{\gamma ,l}(x-\xi ,t)\varphi (\xi
)\,d^{n}\xi +\int_{||x-\xi ||_{p}<p^{-m}}Z_{\gamma ,l}(x-\xi ,t)\varphi (\xi
)\,d^{n}\xi .
\end{equation*}%
Now if $||x||_{p}\geq p^{-m}$, $l>m$, then $Z_{\gamma ,l}(x,t)=Z_{\gamma
}(x,t)$, and 
\begin{equation*}
\frac{1}{\Gamma _{p}^{\left( n\right) }(-\gamma )}%
\int_{||y||_{p}>p^{-l}}||y||_{p}^{-\gamma -n}\bigl(u_{1}(x-y,t)-u_{1}(x,t)%
\bigr)\,dy=
\end{equation*}%
\begin{equation*}
\int_{||x-\xi ||_{p}\geq p^{-m}}Z_{\gamma }(x-\xi ,t)\varphi (\xi )\,d\xi
\end{equation*}%
\begin{equation}
+\int_{||x-\xi ||_{p}<p^{-m}}Z_{\gamma ,l}(x-\xi ,t)\varphi (\xi )\,d\xi ,%
\text{ }  \label{29A}
\end{equation}

for $l>m$. Now using Fubini's theorem, and taking $\lim_{l\rightarrow \infty
}$, we obtain that%
\begin{equation*}
\lim_{l\rightarrow \infty \text{ }}\int_{||x-\xi ||_{p}<p^{-m}}Z_{\gamma
,l}(x-\xi ,t)\varphi (\xi )\,d\xi
\end{equation*}%
\begin{eqnarray*}
&=&\int_{\mathbb{Q}_{p}^{n}}||y||_{p}^{-\gamma -n}\times \\
&&\left( \frac{1}{\Gamma _{p}^{\left( n\right) }(-\gamma )}\int_{||x-\xi
||_{p}<p^{-m}}\bigl(Z(x-\xi -y,t)-Z(x-\xi ,t)\bigr)\varphi (\xi )\,d^{n}\xi
\right) \,d^{n}y
\end{eqnarray*}%
\begin{eqnarray*}
&=&\int_{||x-\xi ||_{p}<p^{-m}}\times \\
&&\left( \frac{1}{\Gamma _{p}^{\left( n\right) }(-\gamma )}\int_{\mathbb{Q}%
_{p}^{n}}||y||_{p}^{-\gamma -n}\bigl(Z(x-\xi -y,t)-Z(x-\xi ,t)\bigr)%
\,d^{n}y\right) \varphi (\xi )\,d^{n}\xi
\end{eqnarray*}%
\begin{equation}
=\int_{||x-\xi ||_{p}<p^{-m}}Z_{\gamma }(x-\xi ,t)\varphi (\xi )\,d\xi
\label{31}
\end{equation}%
Since 
\begin{equation*}
||y||_{p}^{-\gamma -n}\bigl(u_{1}(x-y,t)-u_{1}(x,t)\bigr)
\end{equation*}%
is integrable as function of $y$ (because $u_{1}(x,t)\in \mathfrak{M}%
_{\lambda }$, $\gamma >\lambda $, by Lemma \ref{lemma10}), the result
follows by taking $\lim_{l\rightarrow \infty \text{ }}$ in (\ref{29A}) and
using (\ref{31}).
\end{proof}

\begin{lemma}
\label{claim 4} Let $\lambda <\gamma \leq \alpha $, then 
\begin{equation*}
(D_{T}^{\gamma }u_{2})(x,t)=\int_{0}^{t}\left( \int_{\mathbb{Q}%
_{p}^{n}}(D_{T}^{\gamma }Z)(x-\xi ,t-\tau )f(\xi ,\tau )\,d\xi \right)
\,d\tau .
\end{equation*}
\end{lemma}

\begin{proof}
We set 
\begin{equation*}
u_{2,h}(x,t):=\int_{0}^{t-h}\left( \int\limits_{\mathbb{Q}%
_{p}^{n}}Z(x-y,t-\theta )f(y,\theta )\,d^{n}y\right) \,d\theta .
\end{equation*}%
Then 
\begin{align*}
& \frac{1}{\Gamma _{p}^{\left( n\right) }(-\gamma )}%
\int_{||y||_{p}>p^{-l}}||y||_{p}^{-\gamma -n}\bigl(%
u_{2,h}(x-y,t)-u_{2,h}(x,t)\bigr)\,d^{n}y \\
& =\frac{1}{\Gamma _{p}^{\left( n\right) }(-\gamma )}\int%
\limits_{||y||_{p}>p^{-l}}||y||_{p}^{-\gamma -n}\times \\
& \left( \int_{0}^{t-h}\Bigl(\int\limits_{\mathbb{Q}_{p}^{n}}\bigl(Z(x-y-\xi
,t-\tau )-Z(x-\xi ,t-\tau )\bigr)f(\xi ,\tau )\,d^{n}\xi \Bigl)\,d\tau
\right) \,d^{n}y \\
& =\int_{0}^{t-h}\left( \int\limits_{\mathbb{Q}_{p}^{n}}Z_{\gamma ,l}(x-\xi
,t-\tau )f(\xi ,\tau )\,d^{n}\xi \right) \,d\tau ,
\end{align*}%
with $Z_{\gamma ,l}(x,t)$ as in (\ref{26A}). We now note that 
\begin{equation*}
Z_{\gamma ,l}(x,t)=\int\limits_{\mathbb{Q}_{p}^{n}}\psi (x\cdot \xi
)P_{l}(\xi )e^{-at||\xi ||_{p}^{\alpha }}\,d\xi ,
\end{equation*}%
where 
\begin{equation*}
P_{l}(\xi )=\frac{1}{\Gamma _{p}^{\left( n\right) }(-\gamma )}%
\int\limits_{||y||_{p}>p^{-l}}||y||_{p}^{-\gamma -n}\bigl(\psi (-y\cdot \xi
)-1\bigr)\,dy.
\end{equation*}%
By using \ a similar reasoning to the one used in \cite[pg. 142]{Koch1}, we
have 
\begin{equation*}
|P_{l}(\xi )|\leq \frac{2||\xi ||_{p}^{\gamma }}{|\Gamma _{p}^{\left(
n\right) }(-\gamma )|}\int\limits_{||u||_{p}>1}||u||_{p}^{-\gamma
-n}\,d^{n}u=C||\xi ||_{p}^{\gamma },
\end{equation*}%
whence 
\begin{equation*}
\left\vert Z_{\gamma ,l}(x,t)\right\vert \leq C^{\prime }.
\end{equation*}%
Furthermore, if $||x-\xi ||_{p}\geq p^{-(l-1)}$ then $Z_{\gamma ,l}(x-\xi
,t-\tau )=Z_{\gamma }(x-\xi ,t-\tau )$. Therefore 
\begin{align*}
\int_{0}^{t-h}& \left( \int_{\mathbb{Q}_{p}^{n}}Z_{\gamma ,l}(x-\xi ,t-\tau
)f(\xi ,\tau )\,d^{n}\xi \right) \,d\tau \\
& =\int_{0}^{t-h}\left( \int\limits_{||x-\xi ||_{p}\geq p^{-(l-1)}}Z_{\gamma
}(x-\xi ,t-\tau )f(\xi ,\tau )\,d^{n}\xi \right) \,d\tau \\
& +\int_{0}^{t-h}\left( \int\limits_{||x-\xi ||_{p}<p^{-(l-1)}}Z_{\gamma
,l}(x-\xi ,t-\tau )f(\xi ,\tau )\,d^{n}\xi \right) \,d\tau .
\end{align*}%
By taking $l\rightarrow \infty $ we obtain that 
\begin{equation*}
(D_{T}^{\gamma }u_{2,h})(x,t)=\int_{0}^{t-h}\left( \int_{\mathbb{Q}%
_{p}^{n}}(D_{T}^{\gamma }Z)(x-\xi ,t-\tau )f(\xi ,\tau )\,d^{n}\xi \right)
\,d\tau
\end{equation*}%
\begin{equation*}
=\int_{0}^{t-h}\left( \int_{\mathbb{Q}_{p}^{n}}(D_{T}^{\gamma }Z)(x-\xi
,t-\tau )\left( f(\xi ,\tau )-f\left( x,\tau \right) \right) \,d^{n}\xi
\right) \,d\tau
\end{equation*}%
\begin{equation*}
=\int_{0}^{t-h}\left( \int_{\left\Vert x-\xi \right\Vert
>p^{-l}}(D_{T}^{\gamma }Z)(x-\xi ,t-\tau )\left( f(\xi ,\tau )-f\left(
x,\tau \right) \right) \,d^{n}\xi \right) \,d\tau ,
\end{equation*}%
where $l$ is the exponent of local constancy of $f(\xi ,\tau )$ (c.f.
Corollary \ref{cor2}). Finally, since $u_{2,h}\in \mathcal{M}_{\lambda }$
uniformly in $h$ (c.f. Lemma \ref{lemma2}), by taking $h\rightarrow 0^{+}$
and using the dominated convergence theorem, we have 
\begin{equation*}
(D_{T}^{\gamma }u_{2})(x,t)=\int_{0}^{t}\left( \int_{\mathbb{Q}%
_{p}^{n}}(D_{T}^{\gamma }Z)(x-\xi ,t-\tau )\left( f(\xi ,\tau )-f\left(
x,\tau \right) \right) \,d^{n}\xi \right) \,d\tau .
\end{equation*}
\end{proof}

As a consequence of Lemmas \ref{lemma8}, \ref{claim 3}, and \ref{claim 4},
we obtain the following result.

\begin{proposition}
\label{prop3}%
\begin{align*}
(D_{T}^{\gamma }u)(x,t)& =\int_{\mathbb{Q}_{p}^{n}}(D_{T}^{\gamma }Z)(x-\xi
,t)\varphi (\xi )\,d^{n}\xi \\
& +\int_{0}^{t}\left( \int_{\mathbb{Q}_{p}^{n}}(D_{T}^{\gamma }Z)(x-\xi
,t-\tau )(f(\xi ,\tau )-f(x,\tau ))\,d^{n}\xi \right) \,d\tau ;
\end{align*}%
\begin{align*}
a(D_{T}^{\gamma }u)(x,t)& =-\int_{\mathbb{Q}_{p}^{n}}\frac{\partial Z}{%
\partial t}(x-\xi ,t)\varphi (\xi )\,d^{n}\xi \\
& -\int_{0}^{t}\left( \int_{\mathbb{Q}_{p}^{n}}\frac{\partial Z}{\partial t}%
(x-\xi ,t-\tau )(f(\xi ,\tau )-f(x,\tau ))\,d^{n}\xi \right) \,d\tau ,
\end{align*}%
for $0<\gamma \leq \alpha $.
\end{proposition}

\subsubsection{Proof of Theorem \protect\ref{Thm 1}.}

By Lemma \ref{lemma10}, $u(x,t)\in \mathfrak{M}_{\lambda }$ uniformly with
respect to $t$, and $u(x,t)$ satisfies the initial condition of Theorem \ref%
{Thm 1}. By Propositions \ref{prop2}-\ref{prop3}, $u(x,t)$ is a solution of
Cauchy problem (\ref{Cauchy}).

\subsection{Taibleson Operator and \ Elliptic Pseudo-differential Operators}

For a polynomial $g(x)\in \mathbb{Z}_{p}\left[ x_{1},\ldots ,x_{n}\right] $
we denote by $\overline{g}(x)\in \mathbb{F}_{p}\left[ x_{1},\ldots ,x_{n}%
\right] $ its reduction modulo $p$, i.e., the polynomial obtained by
reducing the coefficients of $g(x)$ modulo $p$. Let $f(x)\in \mathbb{Z}_{p}%
\left[ x_{1},\ldots ,x_{n}\right] $, $f\left( 0\right) =0$, \ be a
non-constant homogeneous polynomial of degree $d$ such that $\overline{f}%
(x)\neq 0$. We say \textit{that }$f(x)$\textit{\ is elliptic modulo }$p$ if 
\begin{equation*}
\left\{ x\in \mathbb{F}_{p}^{n}\mid \overline{f}(x)=0\right\} =\left\{
0\right\} ,
\end{equation*}%
and that $f(x)$\textit{\ is elliptic over }$\mathbb{Q}_{p}$ if 
\begin{equation*}
\left\{ x\in \mathbb{Q}_{p}^{n}\mid f(x)=0\right\} =\left\{ 0\right\} .
\end{equation*}%
Note that if $f$ elliptic modulo $p$, then $f$ is elliptic over $\mathbb{Q}%
_{p}$.

If $I$ is a non-empty subset of $\left\{ 1,\ldots ,n\right\} $, we define $%
f_{I}\left( x\right) $, respectively $\overline{f}_{I}\left( x\right) $, as
the polynomial mapping obtained by restricting $f(x)$ to the set 
\begin{equation*}
T_{I}:=\left\{ x\in \mathbb{Z}_{p}^{n}\mid x_{i}\neq 0\Leftrightarrow i\in
I\right\} ,
\end{equation*}%
respectively, to the set 
\begin{equation*}
\overline{T}_{I}:=\left\{ x\in \mathbb{F}_{p}^{n}\mid x_{i}\neq
0\Leftrightarrow i\in I\right\} .
\end{equation*}

\begin{definition}
Let $f(x)\in \mathbb{Z}_{p}\left[ x_{1},\ldots ,x_{n}\right] $, $f\left(
0\right) =0$, \ be a non-constant homogeneous polynomial of degree $d$ \
with coefficients in $\mathbb{Z}_{p}^{\times }$. We say that $f(x)$ is
strongly elliptic modulo $p$, if for every non-empty subset $I$ of $\left\{
1,\ldots ,n\right\} $, $\overline{f}_{I}\left( x\right) $ is \textit{%
elliptic modulo }$p$.
\end{definition}

\begin{example}
Let $f(x)=x^{2}-\upsilon y^{2}$, with $\upsilon \in \mathbb{Z}_{p}^{\times
}\setminus \left( \mathbb{Z}_{p}^{\times }\right) ^{2}$, where 
\begin{equation*}
\left( \mathbb{Z}_{p}^{\times }\right) ^{2}:=\left\{ x\in \mathbb{Z}%
_{p}^{\times }\mid x=y^{2}\text{, for some }y\in \mathbb{Z}_{p}^{\times
}\right\} .
\end{equation*}%
Then $f(x)$ is strongly elliptic modulo $p$.
\end{example}

\begin{lemma}
\label{lemma15}There are infinitely many strongly elliptic polynomials
modulo $p$.

\begin{proof}
By induction on n, the number of variables. The case $n=1$ is clear. Assume
as induction hypothesis that the result is true for $1\leq n\leq k$, $k\geq
2 $. Let $g\left( x_{1},\ldots ,x_{k}\right) $ be a strongly elliptic
polynomial modulo $p$ of degree $d$. Set any $\upsilon \in \mathbb{Z}%
_{p}^{\times }$ such that $\overline{\upsilon }$ does not have a $l$-th root
in $\mathbb{F}_{p}^{\times }$, for some $l\geq 2$, and $f(x_{1},\ldots
,x_{k+1})=g\left( x_{1},\ldots ,x_{k}\right) ^{l}-\upsilon x_{k+1}^{ld}$.
Then $f(x_{1},\ldots ,x_{k+1})$ is strongly elliptic modulo $p$.
\end{proof}
\end{lemma}

\begin{lemma}
\label{lemma16}Let $f(x)\in \mathbb{Z}_{p}\left[ x_{1},\ldots ,x_{n}\right] $%
, $f\left( 0\right) =0$, \ be a non-constant homogeneous polynomial of
degree $d$ \ with coefficients in $\mathbb{Z}_{p}^{\times }$. If $f(x)$ is
strongly elliptic modulo $p$, then 
\begin{equation}
\left\vert f(x)\right\vert _{p}=\left\Vert x\right\Vert _{p}^{d}\text{, for
any }x\in \mathbb{Q}_{p}^{n}\text{.}  \label{35}
\end{equation}
\end{lemma}

\begin{proof}
We set $A:=\left\{ \left( z_{1},\ldots ,z_{n}\right) \in \mathbb{Z}%
_{p}^{n}\mid \left\vert z_{i}\right\vert _{p}=1\text{, for some }i\right\} $%
.\ Since $f(x)$ is elliptic over $\mathbb{Q}_{p}$, 
\begin{equation*}
\left( \sup_{z\in A}\text{ }\left\vert f(z)\right\vert _{p}\right)
\left\Vert x\right\Vert _{p}^{d}\leq \left\vert f(x)\right\vert _{p}\leq
\left( \inf_{z\in A}\text{ }\left\vert f(z)\right\vert _{p}\right)
\left\Vert x\right\Vert _{p}^{d}\text{,}
\end{equation*}%
(c.f. Lemma 1 in \cite{Z10}). Thus, in order to prove the result it is
sufficient to show that 
\begin{equation*}
\left\vert f\right\vert _{p}\mid _{A}\equiv 1.
\end{equation*}%
Given a non-empty subset $I$ of $\left\{ 1,\ldots ,n\right\} $, we define 
\begin{equation*}
A_{I}=\left\{ x\in A\mid \left\vert x_{i}\right\vert =1\Leftrightarrow i\in
I\right\} .
\end{equation*}%
Then $\cup _{I}A_{I}$ is a partition of $A$ when $I$ runs through all \
non-empty subsets of $\left\{ 1,\ldots ,n\right\} $, \ and to show \ref{35})
it sufficient to prove that%
\begin{equation*}
\left\vert f\right\vert _{p}\mid _{A_{I}}\equiv 1\text{, for every non-empty
subset }I\text{.}
\end{equation*}

Without loss of generality we may assume that $I=\left\{ 1,\ldots ,r\right\} 
$, $1\leq r\leq n$. Thus, if $x\in A_{I}$, then $x_{i}\in \mathbb{Z}%
_{p}^{\times }$, $i=1,\ldots ,r$, and $x_{i}\in p\mathbb{Z}_{p}$, $%
i=r+1,\ldots ,n$, and $\overline{f}(x)=\overline{f}_{I}\left( x\right) \neq
0 $, since\ $f$ is strongly elliptic modulo $p$, therefore $\left\vert
f\right\vert _{p}\mid _{A_{I}}\equiv 1$.
\end{proof}

\section{Markov Processes and Fundamental Solutions}

\begin{theorem}
\label{theo3}The fundamental solution $Z\left( x,t\right) $ is a transition
density of a time- and space-homogeneous non-exploding right continuous
strict Markov process without second kind discontinuities.
\end{theorem}

\begin{proof}
By Proposition \ref{prop1A} (4) the family of operators

\begin{equation*}
\left( \Theta \left( t\right) f\right) \left( x\right) =\tint\limits_{%
\mathbb{Q}_{p}^{n}}Z\left( x-\eta ,t\right) f\left( \eta \right) d\eta
\end{equation*}%
has the semigroup property. We know that $Z\left( x,t\right) >0$ and $\Theta
\left( t\right) $ preserves the function $f\left( x\right) \equiv 1$ (cf.
Proposition \ref{prop1A}). Thus $\Theta \left( t\right) $ is a Markov
semigroup. The requiring properties of the corresponding Markov process
follow from Proposition \ref{prop1A} \ and general theorems of the theory of
Markov processes \cite{D}, see also \cite[Section XVI]{VVZ}.
\end{proof}


\begin{thebibliography}{99}
\bibitem{A-K1} Albeverio S., and Karwoski W. , Diffusion in $p-$adic
numbers. In: K. Ito, H. Hida (Eds.), \ Gaussian Random Fields, pp. 86-99,
1991, World Scientific, Singapore.

\bibitem{A-K2} Albeverio S., and Karwoski W., A random walk on $p-$adics:
the generator and its spectrum, Stochastic Process. Appl. 53 (1994), 1-22.

\bibitem{A-K-S} Albeverio, S.; Khrennikov, A. Yu.; Shelkovich, V. M.,
Harmonic analysis in the $p$-adic Lizorkin spaces: fractional operators,
pseudo-differential equations, $p$-adic wavelets, Tauberian theorems. J.
Fourier Anal. Appl. 12 (2006), no. 4, 393--425.

\bibitem{A-B-K-O} AvetisovA. V., Bikulov A. H., Kozyrev S. V. , and Osipov
V. A , $p-$adic \ models of ultrametric diffusion constrained by
hierarchical energy landscapes, J. Phys. A: Math. Gen. 35 (2002), 177-189.

\bibitem{A-B-O} AvetisovA. V., Bikulov A. H., and OsipovV. A. , $p-$adic
description of characteristic relaxation \ in complex systems, J. Phys. A:
Math. Gen. 36 (2003), 4239-4246.

\bibitem{Bla} Blair A. D. , Adelic path integrals, Rev. Math. Physics
7(1995), 21-49.

on a variety, Invent. Math. 77 (1984), 1--23.

\bibitem{D} Dynkin E. B. , Foundations of the Theory of Markov Processes,
Prentice-Hall, Englewood Cliffs, NJ, 1961.

\bibitem{Ha1} Haran S., Potentials and explicit sums in arithmetic, Invent.
Math. 101(1990), 797-703.

\bibitem{Ha2} Haran S., Analytic potential theory over the $p-$adics, Ann.
Inst. Fourier 43(1993), 905-944.

\bibitem{I} R. S. Ismagilov, On the spectrum of the self-adjoint operator in 
$L_{2}(p)$ where $p$ is a local field; an analog of the Feynman-Kac formula,
Theor. Math. Phys. 89(1991), 1024-1028.

\bibitem{Kh1} Khrennikov A., $p$-adic valued distributions in mathematical
physics, Kluwer, Dordrecht, 1994.

\bibitem{Kh2} Khrennikov \ A., Non-archimedean \ analysis: Quantum
paradoxes, dynamical systems and biological models, Kluwer, Dordrecht, 1997.

\bibitem{Koch1} Kochubei A. N., Pseudodifferential equations and stochastics
over non-Archimedean fields, Marcel Dekker, 2001.

\bibitem{Koch2} Kochubei A. N., Parabolic equations over the field of $p-$%
adic numbers, Math. USSR Izvestiya 39(1992), 1263-1280.

\bibitem{Koch3} Kochubei A. N., Parabolic Pseudodifferential Equations,
Hypersingular Integrals, and Markov Processes, Math. USSR Izvestiya \textbf{%
33} (1989), 233--259.

\bibitem{R-T} Rammal R., and Toulouse G., Ultrametricity for physicists,
Rev. Modern Physics 58 (1986), 765-778.

\bibitem{TA} Taibleson M.H. , Fourier analysis on local fields. Princeton
University Press, Princeton, N.J.; University of Tokyo Press, Tokyo, 1975.

\bibitem{Va} Varadarajan V. S., \ Path integrals \ for a class of $p-$adic
Schr\"{o}dinger equations, Lett. Math. Phys. 39(1997), 97-106.

\bibitem{VVZ} Vladimirov V. S., Volovich I. V., and Zelenov E. I. , \ $p-$%
adic Analysis \ and mathematical physics. Series on Soviet and East European
Mathematics, 1. World Scientific Publishing Co., Inc., River Edge, NJ, 1994.

\bibitem{Z10} Z\'{u}\~{n}iga-Galindo W.A., Parabolic Equations and Markov
Processes Over $p-$adic Fields, \ accepted in Potential Analysis.
\end{thebibliography}
\end{document}